\documentclass{svjour3}                     % onecolumn (standard format)
\smartqed  % flush right qed marks, e.g. at end of proof
\usepackage{graphicx}
\usepackage{amsmath}
\usepackage{amsfonts}
\usepackage{geometry}

\usepackage{natbib}

\ifx\pdfoutput\@undefined\usepackage[usenames,dvips]{color}
\else\usepackage[usenames,dvipsnames]{color}
\IfFileExists{pdfcolmk.sty}{\usepackage{pdfcolmk}}{} 
\fi
\usepackage[plainpages=false,pdfpagelabels,pagebackref=false,naturalnames=true,hyperindex=true,pdftitle={Guiding the Self-organization of Random Boolean Networks},pdfauthor={Carlos Gershenson}]{hyperref}
\hypersetup{colorlinks=true,
urlcolor=Cerulean,linkcolor=BrickRed,citecolor=RoyalBlue,a4paper,
  pdfpagemode=None,
  pdfstartview=FitH}
\usepackage[all]{hypcap}

\newcommand{\avg}[1]{\left< #1 \right>} % for average
 \journalname{Theory in Biosciences}
\begin{document}

\title{Guiding the Self-organization of Random Boolean Networks\thanks{This work was partially supported by SNI membership 47907 of CONACyT, M\'{e}xico.}
}
%\subtitle{Do you have a subtitle?\\ If so, write it here}

%\titlerunning{Short form of title}        % if too long for running head

\author{Carlos Gershenson}
\institute{%C. Gershenson$^{1,2}$  \at
%$^{1}$
 Instituto de Investigaciones en Matem\'aticas Aplicadas y en Sistemas \\
Universidad Nacional Aut\'onoma de M\'exico\\
Ciudad Universitaria, A.P. 20-726\\
01000 M\'exico D.F. M\'exico\\
%Tel. +52 55 56 22 36 19 \
%Fax +52 55 56 22 36 20 \\
\email{cgg@unam.mx} \
\url{http://turing.iimas.unam.mx/~cgg} \\
%$^{2}$
% Centro de Ciencias de la Complejidad \\
%Universidad Nacional Aut\'onoma de M\'exico\\
}

\date{Received: date / Accepted: date}
% The correct dates will be entered by the editor

\maketitle

\begin{abstract}
Random Boolean networks (RBNs) are models of genetic regulatory networks. It is useful to describe RBNs as self-organizing systems to study how changes in the nodes and connections affect the global network dynamics. 
This article reviews eight different methods for guiding the self-organization of RBNs.  
In particular, the article is focussed on guiding RBNs towards the critical dynamical regime, which is near the phase transition between the ordered and dynamical phases. The properties and advantages of the critical regime for life, computation, adaptability, evolvability, and robustness are reviewed. The guidance methods of RBNs  can be used for engineering systems with the features of the critical regime, as well as for studying how natural selection evolved living systems, which are also critical.

% \PACS{PACS code1 \and PACS code2 \and more}
% \subclass{MSC code1 \and MSC code2 \and more}
\end{abstract}

\section{Self-organization and how to guide it}
\label{sec:selforg}

The concept of self-organization originated within cybernetics 
\citep{Ashby1947sos,vonFoerster1960,Ashby1962} and has propagated into almost all scientific disciplines \citep{NicolisPrigogine1977,Luhmann:1995,Turcotte:2002,CamazineEtAl2003,SkarCoveney2003}. Given the broad domains where self-organization can be described, its formal definition is problematic \citep{GershensonHeylighen2003a,Heylighen2003sos,GershensonDCSOS,Prokopenko:2008}. Nevertheless, we can \emph{use} the concept to study a wide variety of phenomena. 

To better understand self-organization, the following notion can be used: \emph{A system described as self-organizing is one in which elements interact, achieving dynamically a global function or behavior} \cite[p. 32]{GershensonDCSOS}. 
In other words, a \emph{global} pattern is produced from \emph{local} interactions.

Examples of self-organizing systems include a cell (molecules interact to produce life), a brain (neurons interact to produce cognition), an insect colony (insects interact to perform collective tasks), flocks, schools, herds (animals interact to coordinate collective behavior), a market (agents interact to define prices), traffic (vehicles interact to determine flow patterns), an ecosystem (species interact to achieve ecological homeostasis), a society (members interact to define social properties such as language, culture, fashion, esthetics, ethics, and politics). In principle, almost any system can be described as self-organizing \citep{Ashby1962,GershensonHeylighen2003a}. If a system has a set of ``preferred" states, i.e. attractors, and we call those states organized, the system will self-organize towards them. It is useless to enter an ontological discussion on self-organization. Rather, the question is: when is it \emph{useful} to \emph{describe} a system as self-organizing? For example, a cell can be described as self-organizing, but also as a Boolean variable (0=dead, 1=alive). Which description is more accurate? It depends on the aim of the description (model). A model cannot be judged independently of the context where it is used.

Self-organization is a useful description when at least two levels of description are present (e.g. molecules and cells, insects and colony) and we are interested in studying the relationship between the descriptions at these two levels (scales). In this way, one can describe how the interactions at the lower level affect the properties at the higher level. When only the interactions at the lower level are defined, the system can \emph{adapt} to novel situations and be \emph{robust} to perturbations, since the precise global behavior is not predefined. Because of this, the properties of self-organizing systems can be exploited in design and engineering \citep{GershensonDCSOS,Watson:2010}.

The balance between self-organization and design is precisely the aim of guided self-organization (GSO) \citep{Prokopenko:2009}. Although it is difficult to define, GSO can be described as \emph{the steering of the self-organizing dynamics of a system towards a desired configuration}. Cybernetics \citep{Wiener1948,Ashby1956} had already a similar aim, although there was a stronger focus on control and communication, as opposed to self-organization and information. 

The dynamics of self-organizing systems lead them to an ``organized" state or configuration. However, there can be several potential configurations available.
The emerging study of GSO explores the constraints and conditions where self-organizing dynamics can be lead to a particular configuration. Similar to several ``synthetic" approaches \citep{Steels1993}, GSO can be useful on the one hand for understanding how natural systems self-organize and on the other hand for building artificial systems capable of self-organization. This paper focuses on both aspects, exploiting the generality of random Boolean networks (RBNs): First, how can evolution guide the self-organization of genetic regulatory networks? Second, how can we manipulate RBNs to guide their self-organization towards a desired regime?

In the next section, random Boolean networks are briefly reviewed. In Section \ref{sec:selforgInRBNs} the self-organization of RBNs is described. Section \ref{sec:guidingSelforgRBNs} mentions eight different ways in which the self-organization of RBNs can be guided. Section \ref{sec:disc} presents a discussion. Conclusions close the paper.

\section{Random Boolean networks}
\label{sec:RBNs}

Random Boolean networks (RBNs) were originally proposed as models of genetic regulatory networks \citep{Kauffman1969,Kauffman1993}. However, their generality has triggered an interest in them beyond their original purpose \citep{AldanaEtAl2003,Gershenson2004c}.

\begin{figure}[htbp]
\begin{center}
\includegraphics{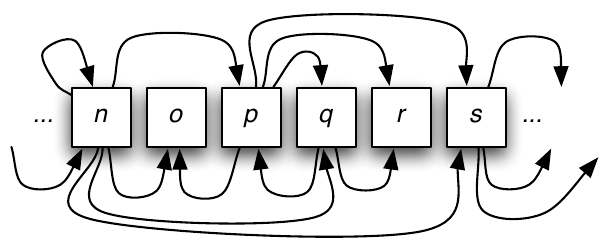}
\caption{Example of a  RBN with connectivity $K=2$, i.e. the state of nodes is determined by the state of two other nodes. Note that all nodes have two inputs, but not necessarily two outputs, e.g. node $n$ affects four other nodes, while node $o$ does not affect any other node.}
\label{fig:topoRBN}
\end{center}
\end{figure}

\begin{figure}[htbp]
\begin{center}
\includegraphics[width=7cm]{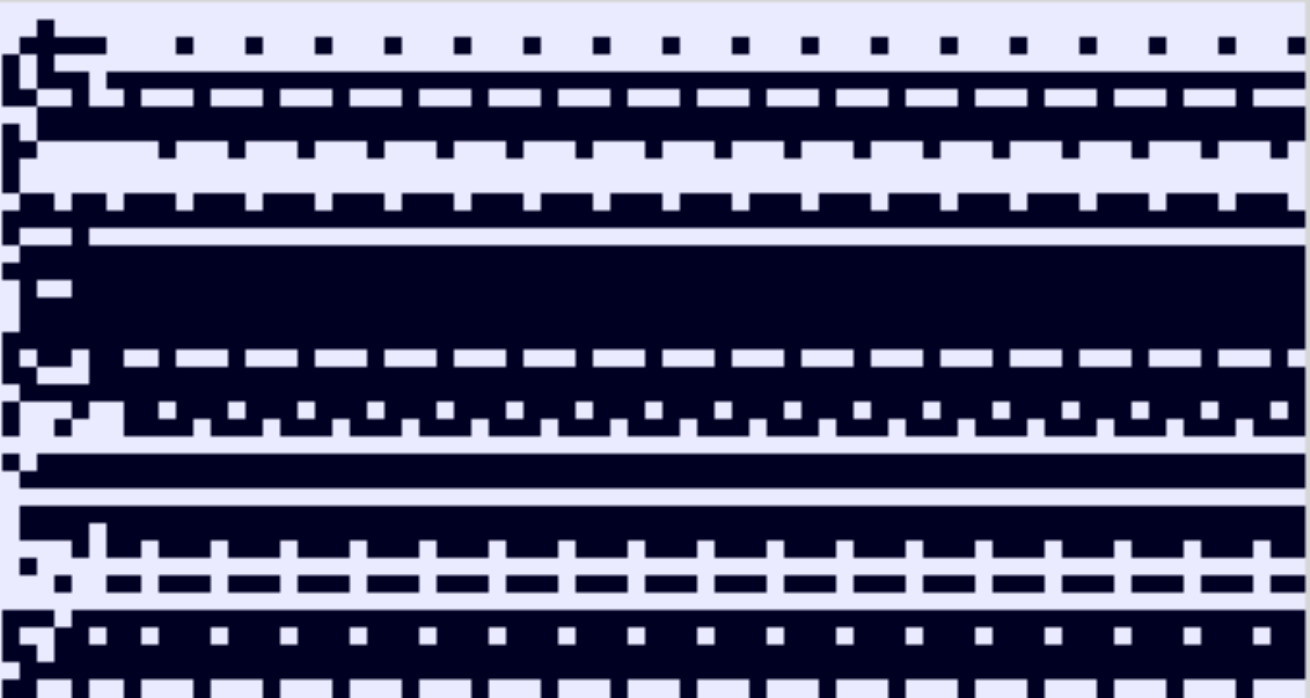}
\caption{Temporal evolution of a RBN with $N=40, K=2$ for a random initial state. Dark squares represent 0 and light squares represent 1. Time flows to the right, i.e. columns represent states of the network at a particular time, while a row represents the temporal evolution of the state of a node. Taken from RBNLab \citep{RBNLab}.}
\label{fig:dynamicsRBN-N40K2}
\end{center}
\end{figure}

A RBN consists of $N$ nodes linked by $K$ connections each. Nodes are Boolean, i.e. their state is either ``on" (1) or ``off" (0). The state of a node at time $t+1$ depends on the states of its $K$ inputs at time $t$ by means of a Boolean function. Connections and functions are chosen randomly when the RBN is generated and remain fixed during its temporal evolution. The randomly generated Boolean functions can be represented as lookup tables that represent all possible $2^K$ combinations of input states. Fig. \ref{fig:topoRBN} shows an example of a part of a RBN, where every node has exactly two inputs, i.e. $K=2$. Table \ref{tab:rules} shows an arbitrary lookup table to update the state of one of the nodes. The dynamics of a RBN with $N=40, K=2$ can be seen in Fig. \ref{fig:dynamicsRBN-N40K2}.

\begin{table}[htbp]
\caption{Lookup table to update node $o$ depending on the state of nodes $n$ and $p$. Lookup tables include all possible combinations of inputs, i.e. $2^K$ rows. Different nodes will have different lookup tables, i.e. Boolean functions.}
\label{tab:rules} \begin{tabular}{lll}
\hline\noalign{\smallskip}
n(t) & p(t) & o(t+1)  \\
\noalign{\smallskip}\hline\noalign{\smallskip}
0 & 0 & 1 \\
0 & 1 & 0 \\
1 & 0 & 0 \\
1 & 1 & 1 \\
\noalign{\smallskip}\hline
\end{tabular}
\end{table}

\begin{figure}[htbp]
\begin{center}
\includegraphics{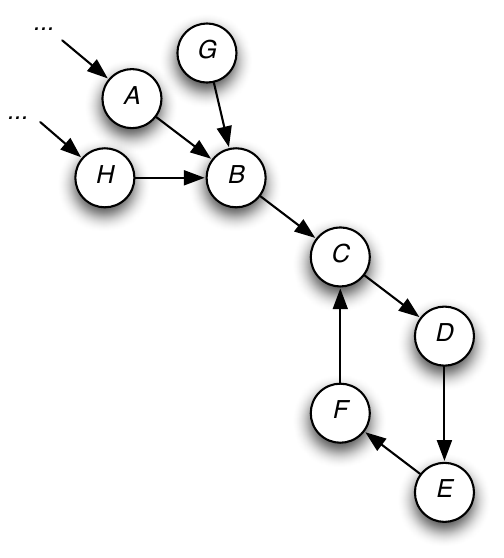}
\caption{Example of state transtions. $B$ is a successor state of $A$ and a predecessor of $C$. States can have many predecessors (e.g. $B$), but only one successor. $G$ is a Garden of Eden state since it has no predecessors. The attractor $C\rightarrow D\rightarrow E \rightarrow F \rightarrow C$ has a period four.}
\label{fig:statesRBN}
\end{center}
\end{figure}

Since RBNs are finite (they have $2^N$ possible states) and deterministic, eventually a state will be revisited. Then the network will have reached an \emph{attractor}. The number of states in an attractor determines the \emph{period} of the attractor. Point attractors have period one (a single state), while cyclic attractors have periods greater than one (multiple states, e.g. four in Fig. \ref{fig:dynamicsRBN-N40K2}). A RBN can have one or more attractors. The set of states visited until an attractor is reached is called a transient. The set of states leading to an attractor form its \emph{basin}. The basins of different attractors divide the state space. RBNs are \emph{dissipative}, i.e. many states can flow into a single state (one state can have several predecessors), but from one state the transition is deterministic towards a single state (one state can have only one successor). 
The number of predecessors is also called in-degree. States without a predecessor are called ``Garden of Eden" (GoE) states (in-degree=0), since they can only be reached from an initial condition. 
Fig. \ref{fig:statesRBN} illustrates the concepts presented above.

Note the difference between the topological network of a RBN (e.g. Fig. \ref{fig:topoRBN})---which represents how the states of nodes affect each other---and its state network (e.g. Fig. \ref{fig:statesRBN})---which represents the transitions of the whole state space. In the state network, each node represents a state of the network, i.e. there are $2^N$ nodes in the state network, while RBN nodes are represented in the topological network, i.e. there are $N$ nodes in the topological network. One of the main topics of RBN research is to understand how changes in the topological network (lower scale) affect the state network (dynamics of higher scale), which is far from obvious.

RBNs are a type of discrete dynamical network \citep{Wuensche1998}, i.e. space, time, and states are discrete. RBNs are generalizations of Boolean cellular automata \citep{vonNeumann1966,Wolfram1986,Wolfram:2002}, where the states of cells are determined by $K$ neighbors, i.e. not chosen randomly, and all cells are updated using the same Boolean function \citep{Gershenson2002e}.

\section{Self-organization in random Boolean networks}
\label{sec:selforgInRBNs}

Random Boolean networks can be described as self-organizing systems simply because they have attractors. If we describe the attractors as ``organized", then the dynamics self-organize towards them \citep{Ashby1962}. Still, a better argument in favor of this description is that we are interested in understanding how the interactions between nodes (lower scale) affect the network dynamics and properties (higher scale). The concept of self-organization allows us to describe and relate both scales and their interactions under the same framework.

The self-organization of RBNs can also be interpreted in terms of \emph{complexity reduction}. For example, the human genome has approximately 25,000 genes. Thus, in principle, each cell could be in one of the $2^{25,000}$ possible states of that network. This is much more than the estimated number of elementary particles arising from the Big Bang. However, there are only about 300 cell types (attractors \citep{Kauffman1993,HuangIngber2000}), i.e. cells self-organize towards a very limited fraction of all possible states. The main question addressed by these review paper is: in which ways can the self-organization of random Boolean networks be guided?

Before presenting multiple answers to that question, it is convenient to understand the different dynamical behaviors that RBNs can have \citep{Wuensche1998,Gershenson2004c}. There are two dynamical phases: \emph{ordered} and \emph{chaotic}. The phase transition is characterized by its \emph{criticality} and is also known as the ``edge of chaos"  \citep{Kauffman1993}. 

In the ordered phase, most nodes do not change their state, i.e. they are static. RBNs are robust in this phase, i.e. damage does not spread through the network, since most nodes do not change. Also, similar states tend to converge to the same attractor. On average, states have many predecessors, which leads to a high convergence (many states go to few states), short transient times, and a high density of Garden of Eden states, i.e. the percentage of all states without a predecessor is high.

In the chaotic phase, most nodes are changing their state. Thus, damage spreads through chaotic networks. Therefore, RBNs are fragile in this phase. Similar states tend to diverge towards different attractors. On average, states have few predecessors, which leads to a low convergence, very long average transient times, and a relatively lower density of Garden of Eden states.

In the critical regime, i.e. close to the transition between the ordered and chaotic phases, the extremes of both phases are balanced: some nodes change and some are static. Therefore, damage or changes can spread, but not necessarily through all of the network. Similar states tend to lie in trajectories that neither converge nor diverge in state space \cite[p. 171]{Kauffman2000}. Few nodes have many predecessors, while many nodes have few predecessors. Actually, the in-degree distribution approximates a power law \citep{Wuensche1998}. There is medium convergence. It has recently been found that RBNs near the critical regime maximize information storage and coherent information transfer \citep{Lizier:2008}, as well as maximize Fisher information \citep{Wang:2010}. 

It has been argued that computation and life should occur at the edge of chaos \citep{Langton1990,Kauffman1993,Crutchfield:1994,Kauffman2000}. Even when criticality seems not to be a necessary condition for complexity \citep{Mitchell:1993}, there is experimental evidence that the genetic networks of organisms from at least four kingdoms are near or within the critical regime \citep{Balleza:2008}.
The tendency towards criticality can be explained because of the following: On the one hand, ordered dynamics produce stability (robustness) which is desirable for preserving information (memory). However, a static system is not able to compute or adapt. On the other hand, chaotic dynamics give variability (exploration), which is necessary for computation and adaptation. Still, there is too much change within the chaotic phase to preserve information. A balance is reached in the critical regime, where the advantages of both phases can coexist: there can be enough stability and robustness to preserve information and enough variability to compute and explore. For this reason, it becomes a relevant question to ask how can we guide the self-organization of RBNs towards the critical regime. Being general models, the answers will give us information on how to achieve the same guidance within particular systems.

\section{Guiding the self-organization of random Boolean networks}
\label{sec:guidingSelforgRBNs}

The criticality of RBNs can depend on many different factors. These factors can be exploited---by 
engineers or by natural selection---to guide the self-organization of RBNs and similar systems towards the critical regime. 
\subsection{$p$} 

One of the most obvious factors affecting the dynamics of a RBN is the probability $p$ of having ones on 
the last column of lookup tables \citep{DerridaPomeau1986}. If $p = 1$, then all values in lookup tables 
will be one, so actually there will be no dynamics: all nodes will have a state of one after one iteration, 
independently on the initial state. The same case but with zero occurs for $p = 0$. When $p = 0.5$ there is 
the maximum variability possible in the lookup tables, i.e. no bias. As $p$ approaches 0 or 1, RBNs tend 
to be more static, i.e. in the ordered regime.

\subsection{$K$}
 
One of the most important factors determining the dynamical phase of RBNs is the connectivity K 
\citep{DerridaPomeau1986,LuqueSole2000}. For $p = 0.5$, the ordered phase is found when 
$K < 2$, the chaotic phase occurs for $K > 2$, while the critical regime lies at the phase transition, i.e. $K = 2$. As $p$ tends towards one or zero, the phase transition moves towards greater values of $K$.

If we focus on a single node $i$ and calculate the probability that damaging its state (be it 0 or 1) will percolate changes through the network, then it is clear that the probability will increase with the connectivity $K$. We can choose a node $j$ from one of the nodes that $i$ can affect. There is a probability $p$ that $j$ will be 1, and thus a damage in $i$ will modify $j$ with a probability $1-p$. The complementary case is the same. Now, for $K$ nodes, we can expect that at least one change will occur if $\avg{K}2p(1-p)\geq 1$ \citep{LuqueSole1997}, i.e. chaos. Generalizing, the critical connectivity becomes \citep{DerridaPomeau1986}:

\begin{equation}
\avg{K_c} = \frac{1}{2p(1-p)}
\end{equation}

\subsection{Canalizing functions}

A canalizing function  \citep{Kauffman1969,Stauffer1987,Szejka:2007} is one in which at least one of the inputs has one value that is able to determine the value of the output of the function, regardless of the other inputs \citep{Shmulevich:2004}. In other words, the non-canalizing inputs of canalizing functions are not relevant. A different type of canalization is considered when a particular value of an input determines the output of a function, while for other(s) values of that input, the output of the function is determined by other inputs.

Independently of the particular type of canalization, if there is a bias favoring canalizing functions, the phase transition will move towards greater values of $K$. This is possible because in practice non-canalizing inputs are ``ficticious", i.e. removing them does not affect the state space nor the dynamics of a RBN.

For more than 150 transcriptional systems, it has been found that there is a strong canalizing bias \citep{Harris:2002}. Systems have $\avg{K}$=3,4, or 5 for most cases, and few with even  $\avg{K}$=7,8, or 9 and still fall within the ordered phase.

It has been shown that RBNs with nested canalizing functions have stable (ordered) dynamics, approaching criticality for low values of $K$ \citep{Kauffman:2003,Kauffman:2004}.

Schemata can be used to describe canalization in RBNs in a concise way \citep{Marques-Pita:2008}.

\subsection{Silencing}

During normal cell life, genes can be ``silenced", i.e. switched off by different mechanisms. 
Silencing is also a method used for perturbing genetic regulatory networks. The RBN equivalent of silencing would be to fix the value of a subset of nodes, independently on the state of their inputs. 

It has been shown that even chaotic networks can be forced into regular behavior when a subset of nodes is not responsive to the internal network dynamics \citep{LuqueSole1997b}. It is straightforward to assume that if a higher percentage of nodes remains fixed, the dynamics will be more stable.

Some studies of silencing on RBNs have been made, e.g. \citet{Serra:2004}, although the precise relationship between silencing and criticality still remains to be studied.

\subsection{Topology}
Until recently, RBN studies used either a homogeneous topology ($K$ is the same for all nodes) or a normal topology (there is an average $\avg{K}$ inputs per node). This implies a uniform input distribution and no regularity in the wiring of nodes. However, the particular topology can have drastic effects on the properties of RBNs. 

\subsubsection{Link distribution}

On the one hand, topologies with more uniform rank distributions, as those used commonly, exhibit more and longer attractors, but with less correlation in their expression patterns. On the other hand, skewed topologies exhibit less and shorter attractors, but with more correlations (entropy and mutual information) \citep{OosawaSavageau2002}. A balance between these two extremes is achieved with scale-free topologies, i.e. few nodes have many inputs, while many nodes have few inputs. This advantageous balance can be used to explain why most natural networks have a scale-free topology \citep{AldanaCluzel2003,Oikonomou:2006}.

RBNs with a scale-free topology \citep{Aldana2003} have been found to expand the advantages of the critical regime into the ordered phase, since well-connected elements can lead to the propagation of changes, i.e. adaptability even when the average connectivity would imply a static regime. It can be said that a scale-free topology expands the range of the critical regime. It has also been found that for RBNs with a scale-free distribution of outputs the average number of attractors is independent of networks size for more than three orders of magnitude \citep{Serra:2003}.

\subsubsection{Link regularity}

Classical RBNs have the same probability of linking any node to any other node, i.e. they are random networks. For the same input distributions, the opposite extreme are regular networks, i.e. where nodes are connected to their neighbors, in a cellular automata fashion. The balance between those two extremes is achieved with small-world networks \citep{Watts:1998,Bullock:2010}: many connections to neighbors and a few long-range connections lead to reduced average path lengths (common of random networks), while maintaining a high clustering coefficient (common of regular networks) \citep{Watts:1998}.

In RBNs, a small world topology maximizes information transfer \citep{Lizier:2010}, which is an indicator of critical dynamics.

\subsection{Modularity}

It is well know that modularity is a prevalent property of natural systems  \citep{Callebaut:2005} and a desired feature of artificial systems \citep{Simon1996}. Modularity is a property difficult to define precisely, but we can agree that as system is modular if it is composed by modules, i.e. the interactions within modules are more relevant than those between modules.
Modules offer a level of organization that promotes at the same time robustness and evolvability \citep{Wagner2005}. On the one hand, damage within one module usually does not propagate through the whole system (robustness). On the other hand, useful changes can be exploited to find new configurations without affecting the functionality of other modules.

In the context of RBNs, initial explorations suggest that topological modules broaden the range of the critical regime towards higher connectivities \citep{BalpoGershenson:2010}, i.e. a modular structure promotes critical dynamics within the theoretical chaotic phase. This is because even with high average connectivities, changes within one node have a low probability of propagating to other modules if there are few intermodular connections, in comparison with a non-modular network. These studies are related to topological modularity. One could argue that functional modularity is related to topological modularity \citep{Calabretta:1998}, although this precise relationship remains to be studied.

\subsection{Redundancy}

Redundancy consists of having more than one copy of an element type. Duplication combined with mutation is a usual mechanism for the creation of genes in eukaryotes  \citep{FernandezSole2003}. When there are several copies of an element type, changes or damage can occur to one element while others continue to function.

For RBNs \citep{GershensonEtAl2006}, redundancy of links is not useful, since redundant links are fictitious inputs, i.e. they do not affect the state space. However, a redundancy of nodes prevents mutations from propagating through the network. Thus, redundant nodes increase neutrality \citep{Kimura1983,vanNimwegen:1999,Munteanu:2008}, i.e. can ``smoothen" rough landscapes. This is an advantage for robustness and for evolvability, for the same reasons as the ones discussed concerning modularity, even when redundancy is a different mechanism from modularity, although both can potentially be combined.

\subsection{Degeneracy}

Degeneracy---also known as distributed robustness---is defined as the ability of elements that are structurally different to perform the same function or yield the same output \citep{Edelman:2001,FernandezSole2003}. As modularity, it also widespread in biological systems and a promotor of robustness and evolvability. There is evidence that in genetic networks distributed robustness is equally or more important for mutational robustness  \citep{Wagner2004,Wagner2005} and for evolvability \citep{Whitacre:2010} than gene redundancy. 

To date, particular studies of degeneracy on RBNs are lacking. Nevertheless, it could be speculated that degeneracy should promote critical dynamics, even when this still remains to be explored.

\section{Discussion}
\label{sec:disc}

In the previous section, a non-exhaustive list of factors that can be used to guide the self-organization of RBNs towards the critical regime was presented. Two categories of methods can be identified for guiding the self-organization towards criticality: moving the phase transition (with $p$, $K$, canalizing functions, or silencing) or broadening the critical regime (with balanced topologies, modularity, redundancy, or degeneracy). The first category seems to be directed more at the node functionality and lookup tables, while the second category seems to be directed more at the connectivity between nodes.

It can be speculated that natural selection can exploit these and probably other methods to guide the self-organization of genetic regulatory networks towards the critical regime. There is evidence that some of these methods are exploited by natural selection, but further studies are required to understand better the mechanisms, their constraints, how they are related, and which ones have been actually employed and to what degree by natural selection.
In a similar way, engineers can guide the self-organization of RBNs and related systems using these and other methods. The purpose of designing artificial systems with criticality is to provide them with the advantages of this dynamical regime, that living systems have used to survive in an unpredictable environment.

Concerning the methods that move the phase transition, if a RBN is in the ordered phase, one or several of the following can be done:

\begin{itemize}
\item Adjust $p$ towards a value of 0.5. This will increase variety in lookup tables, and thus increase the number of nodes that change. In other words, dynamics will be promoted.
\item Increase the connectivity $K$. More connections also promote richer dynamics.
\item Decrease the number of canalizing functions (if any). Canalizing functions imply that some connections play no role on the dynamics. If these connections are changed, i.e. they become functional, dynamics will be richer. 
\end{itemize}

If the RBN is in the chaotic phase, complementary measures can be taken:
\begin{itemize}
\item Adjust $p$ farther from a value of 0.5. This will increase homogeneity in lookup tables, and less nodes will be changing. This will decrease the damage sensitivity of the network, i.e. the dynamics will be less chaotic and more stable.
\item Decrease the connectivity $K$. Less connections promote stability.
\item Increase the number of canalizing functions. Canalizing functions reduce the effect of having several inputs, i.e. a high connectivity $K$, so changes cannot propagate as easily as with no canalizing functions. 
\item Silence some nodes by fixing their state independently of their inputs.
\end{itemize}

Concerning the methods that broaden the critical regime, even when they are different, all of them can be exploited to guide the self-organization of RBNs towards the critical regime:

\begin{itemize}
\item Promote a scale-free topology. When few nodes have many connections and most nodes have few connections, the desirable properties of the critical regime extend beyond the extremely narrow space of all possible networks that lies precisely on the phase transition (e.g. $K=2$, $p=0.5$). This ``criticality enhancement" is possible because most nodes are stable, but the nodes with several connections (hubs) can trigger rich dynamics. In this way, changes can propagate through the RBN in a constrained fashion.
\item Promote a small world topology. Having a high clustering coefficient and a low average path length balances several advantages that lead to critical dynamics \citep{Lizier:2010}.
\item Promote modularity. Modules make it difficult for damage to spread through all of the network, even if the local connectivity (within a module) is high. In this way, chaotic dynamics can be constrained within modules. This prevents avalanches, where change to one node might cause drastic changes in a large part of the network. Still, modularity allows for information flow between modules.
\item Promote redundancy. Having more than one copy of a node (or module) implies that a change on that node (or module) will not propagate through the network, since the redundant element(s) can perform the same function. Apart from smoothening rough fitness landscapes \citep{GershensonEtAl2006},
it has been noted that redundancy is a useful feature in evolvable hardware \citep{Thompson1998}. 
\item Promote degeneracy. The effect of degeneracy is similar to that of redundacy, but acting at a functional level. Different components of a system perform the same function. In certain conditions, degeneracy might be advantageous over redundancy, e.g. when a change affects all copies of the same node (or module). Nevertheless, redundancy seems to be useful for exploration via duplication and mutation.
\end{itemize}

Yet another way of guiding the self-organization of RBNs towards the critical regime would be to promote certain properties as a part of the fitness function of an evolutionary algorithm. For example, one could evolve critical RBNs trying to maximize input entropy variance \citep{Wuensche1999}, information storage, information transfer \citep{Lizier:2008}, and/or Fisher information  \citep{Wang:2010}. Another criterion could be to try to approximate Lyapunov exponents to zero \citep{LuqueSole2000}. All of these properties characterize the critical regime. Thus, they can be used as a guidance of the evolutionary search. Nevertheless, it should be noted that guiding the self-organization of RBNs with a fitness function that promotes criticality is not useful to explain how natural systems evolved their criticality. Still, they are a valid approach for engineering critical systems.

It has been noted that the updating scheme can affect the behavior of RBNs \citep{HarveyBossomaier1997,Gershenson2002e}. However, it does not affect the transition between the ordered and chaotic phases \citep{Gershenson2004b}. Still, within the chaotic phase, random mutations have less effect when the updating is non-deterministic. However, this is because basins of attraction merge with a less-deterministic updating scheme \citep{GershensonEtAl2003a}, i.e. there are less attractors. On the one hand, non-determinism reduces sensitivity to random mutations. On the other hand, non-determinism reduces the functionality of RBNs. Having this tradeoff, and many different possible updating schemes \citep{Giacobini:2006,Darabos:2009}, it is difficult to argue in favor of any updating scheme in the context of this paper.

\subsection{Why criticality?}

Some of the advantages of the critical regime were already presented, namely the balance between stability and variance that are requirements for life \citep{Kauffman1993,Kauffman2000} and computation \citep{Langton1990,Crutchfield:1994}. In addition, the critical regime is also advantageous for adaptability, evolvability, and robustness.

\emph{Adaptability} can be understood as the ability of a system to produce advantageous changes in response to an environmental or internal state that will help the system to fulfill its goals \citep{GershensonDCSOS}. Suppose that a system that is modelled by a RBN (such as a genetic regulatory network) is situated in an unpredictable environment. It is desirable that the system will be able to adapt to changes in the environment. Does criticality increase adaptability? Not by itself, but it is useful. An ordered RBN will not be able to adapt so easily, because most changes will have no effect on the dynamics, so there will be no response to the environmental change. A chaotic RBN will pose the opposite difficulty: most changes will have a strong effect on the dynamics. Thus, it is highly probable that some of the functionality of the system will be lost. A balance is achieved by a critical RBN: changes can have an effect on the dynamics, but their propagation can be constrained, preserving most of the functionality of the system. 

\emph{Evolvability} is the ability of random variations to sometimes produce improvement \citep{Wagner:1996}. It can be seen as a particular type of adaptability, where changes occur from generation to generation.
RBN evolution has already been explored \citep{Stern1999,LemkeEtAl2001}. For the same reasons as those exposed for adaptability, evolvability will be higher for critical RBNs. Within an evolutionary context, however, it is also important to mention the advantages of criticality to scalability \citep{Simon1996}, i.e. the ability to acquire novel functionalities. Since ordered RBNs have restricted dynamics and interactions, they cannot integrate novel functions too easily. Chaotic RBNs are also problematic, since novel functions most probably change the existing functionality. Critical RBNs are scalable, since novel functions can be integrated without altering existing functionalities. Intuitively, modularity promotes scalability in the most straightforward way, although other methods---i.e. critical RBNs without modularity---also allow for scalability.

A system is \emph{robust} if it continues  to function in the face of perturbations  \citep{Wagner2005,Jen2005}. Robustness is a desirable property to complement adaptability and evolvability, since changes in the environment (perturbations) can damage or destroy a system before it can adapt or evolve. It is clear that evolution is not possible without robustness. 
Chaotic RBNs are not robust, since small perturbations produce drastic changes. Ordered RBNs are very robust, since they can resist most perturbations without producing changes. And when changes are produced, these do not propagate. However, ordered RBNs do not offer rich dynamics. Critical RBNs offer both advantages: rich dynamics and robustness (although not as high as the one of ordered RBNs. Note that the most robust RBNs are those without dynamics, e.g. with $p=1$.). 

Topology and modularity seem to be more relevant for adaptability and evolvability, while redundancy and degeneracy seem to be more relevant for robustness. However, evolvability and robustness are interrelated properties \citep{Yu:2001,Ebner:2002,Wagner2005}, both of them desirable in natural and artificial systems.

It has been argued by \citet{Riegler:2008} that canalization is indispensable for the evolvability of complexity.

During the development of organisms, it seems that the ``perfect" balance between adaptability and robustness changes with age, i.e. embryos are more plastic than adults \citep[ch. 13]{Neuman:2008}. Silencing seems to be a method used by organisms to tune this tradeoff, although other methods might be involved in this process as well.

\section{Conclusions}
\label{sec:conc}

This paper described random Boolean networks (RBNs) as self-organizing systems. Given the advantages of the critical regime of RBNs, different methods to guide the self-organization of RBNs towards criticality were reviewed.

One can ask: which comes first, criticality or some of the methods that promote it? Do they always come hand in hand? It seems not, since criticality can be present without canalizing functions, modularity, redundancy, degeneracy, or scale free topologies. However, these properties facilitate (guide) criticality. Therefore, there is a selective pressure in favor of these properties. Which are the pressures that have actually guided genetic regulatory networks towards criticality \citep{Balleza:2008} is an open question. Which methods of the ones presented have actually been exploited by natural selection is another open question, although there is evidence of several of them at play. Yet another open question is how are the different methods related. This question actually comprises a whole set of questions, relevant for networks in general, e.g. how are scale-free, small world, and modular topologies related? Is there an advantage of having combinations of them, e.g. a scale-free modular topology, over only one of them? Do Apollonian networks \citep{Andrade:2005} offer a ``maximal" criticality? What would be theoretically the maximal criticality possible for a given family of RBNs? When is redundancy or degeneracy preferable? What are the differences and advantages of critical RBNs produced with one or several of the presented methods? How are different methods related to adaptability, evolvability, and robustness? What is the proper balance between evolvability and robustness?

This long list of relevant questions, which could easily continue growing, should motivate researchers to continue exploring RBNs, their self-organization, and methods for guiding it. The answers should be relevant for the scientific study of networks, artificial life, and engineering.

\begin{acknowledgements}
I should like to thank the guest editors of this special issue and anonymous referees for useful comments.
%This work was partially supported by SNI membership 47907 of CONACyT, M\'{e}xico.
\end{acknowledgements}

% BibTeX users please use one of
\bibliographystyle{spbasic}      % basic style, author-year citations
\bibliography{carlos,RBN,sos,complex,COG,evolution,information}

\end{document}